\documentclass[12pt]{article}
\pdfoutput=1
\usepackage{amsmath}
\usepackage{graphics,graphicx,xcolor}
\def\be{\begin{equation}}
\def\ee{\end{equation}}
\def\arr{\begin{array}{rll}}
\def\ea{\end{array}}
\def\bea{\begin{eqnarray}}
\def\eea{\end{eqnarray}}

\def\N2{$N{=}2$}

\def\>{\rangle}
\def\<{\langle}
\def\+{\dagger}
\def\={\ =\ }

\begin{document}
\renewcommand{\thefootnote}{\fnsymbol{footnote}}
\begin{titlepage}
\setcounter{page}{0}
\vskip 1cm
\begin{center}
{\LARGE\bf  Rational Ruijsenaars--Schneider model}\\
\vskip 0.4cm
{\LARGE\bf   with cosmological constant}\\
\vskip 1cm
$
\textrm{\Large Anton Galajinsky \ }
$
\vskip 0.7cm
{\it
Tomsk Polytechnic University, 634050 Tomsk, Lenin Ave. 30, Russia} \\
\vskip 0.3cm
{\it
Tomsk State University of Control Systems and Radioelectronics,\\
Lenin ave. 40, 634050 Tomsk, Russia
} \\

\vskip 0.2cm
{e-mail: galajin@tpu.ru}
\vskip 0.5cm
\end{center}
\vskip 1cm
\begin{abstract} \noindent
The Ruijsenaars--Schneider models are integrable dynamical realizations of the Poincar\'e group in 
$1+1$ dimensions, which reduce to the Calogero and Sutherland systems in the nonrelativistic limit.
In this work, a possibility to construct a one--parameter deformation of 
the Ruijsenaars--Schneider models by uplifting the Poincar\'e algebra in $1+1$ dimensions 
to the anti de Sitter algebra is studied. It is shown that amendments including a cosmological constant 
are feasible for the rational variant, while the hyperbolic and trigonometric systems are ruled out
by our analysis. The issue of integrability of the deformed rational model is discussed in some detail.
A complete proof of integrability remains a challenge.
\end{abstract}


\end{titlepage}

\renewcommand{\thefootnote}{\arabic{footnote}}
\setcounter{footnote}0

\noindent
{\bf 1. Introduction}\\

The Ruijsenaars-Schneider models \cite{RS,RS1} comprise interesting examples 
of integrable many--body systems, equations of motion of which
involve particle velocities. They furnish dynamical realizations of the
Poincar\'e group in $1+1$ dimensions, which includes translations in temporal and 
spatial directions and the Lorentz boost, and reduce to the celebrated 
Calogero and Sutherland models \cite{C,S} 
in the nonrelativistic limit. By this reason, the former are conventionally regarded 
as relativistic generalizations of the latter. The relativistic systems of such a type 
proved relevant for 
various physical applications including dualities (for a review and further references see \cite{FGNR}).

As is well known, the Poincar\'e algebra can be regarded as a contraction of the (anti) de Sitter 
algebra in which a cosmological constant tends to zero \cite{BL}. 
A natural question arises whether the analysis in \cite{RS} can be extended 
so as to include a cosmological constant. 

The goal of this work is to study a possibility to construct a one--parameter deformation of 
the Ruijsenaars--Schneider models by uplifting the Poincar\'e algebra in $1+1$ dimensions 
to the anti de Sitter algebra. As shown below, amendments including a cosmological constant 
are feasible for the rational variant, while the hyperbolic and trigonometric systems are ruled out
by our analysis. 

The work is organized as follows. In the next section, the original construction in \cite{RS}
is briefly reviewed.  

In Sect. 3, starting 
with the anti de Sitter algebra, which is a one--parameter deformation 
of the Poincar\'e algebra in $1+1$ dimensions, and properly modifying 
the generators which furnish a dynamical realization, two functional equations are obtained, 
which determine interaction potential. The first equation coincides with that in \cite{RS}. 
The second restriction is shown to be compatible with the rational model but it rules out
the trigonometric and hyperbolic variants. Because the anti de Sitter algebra in $1+1$ dimensions is isomorphic to
the conformal algebra $so(2,1)$, the generalized rational model in this work 
provides a new dynamical realization of the conformal group $SO(2,1)$ within the framework of many--body 
mechanics in one dimension. 

In Sect. 4, equations of motion of the rational 
Ruijsenaars--Schneider model with a cosmological constant are constructed. 
The derivation relies upon specific subsidiary functions, which via the Poisson bracket generate 
interaction potential of interest. Similar formalism has previously proved useful for 
constructing supersymmetric 
extensions \cite{AG1}. Like in the nonrelativistic case \cite{GP}, the presence of a cosmological constant results 
in an effective confining potential which renders particles motion (quasi)periodic.

The issue of integrability is discussed in Sect. 5. Constants of motion 
characterizing the three--body case are explicitly constructed. 

We summarize our results and discuss 
possible further developments in the concluding Sect. 6.

Throughout the paper, summation signs are
written out explicitly, i.e. no summation over repeated indices is understood.  

\vspace{0.5cm}

\noindent
{\bf 2. The Ruijsenaars--Schneider models}\\

The Ruijsenaars--Schneider models were originally introduced in \cite{RS} as dynamical realizations of
the Poincar\'e group in $1+1$ dimensions (for a review see \cite{RS1}). The key feature 
of the construction is that the angle $\varphi$ which specifies the Lorentz transformation 
in two--dimensional spacetime
\be\label{tr1}
\left(\begin{array}{cccc}
ct'  \\
x' \\
\end{array}
\right)=\left(
\begin{array}{cccc}
\cosh{\varphi} & \sinh{\varphi}  \\
\vspace{0.2cm}
\sinh{\varphi} & \cosh{\varphi} \\
\end{array}
\right) \left(\begin{array}{cccc}
ct  \\
x \\
\end{array}
\right)
\ee
is represented as the ratio of the rapidity $\theta$ and 
the speed of light $c$
\be\label{vp}
\varphi=\frac{\theta}{c}
\ee
and $\theta$ is taken to be a dynamical variable
instead of the velocity $v$ used within 
the special relativity. 

The well known relation
\be
\tanh{\frac{\theta}{c} }=-\frac{v}{c}
\nonumber
\ee
links (\ref{tr1}) to the conventional form of the Lorentz boost
\be\label{tr2}
t'=\frac{t- \frac{v}{c^2} x}{\sqrt{1-{\left(\frac{v}{c} \right)}^2}}, \qquad 
x'=\frac{x- v t}{\sqrt{1-{\left(\frac{v}{c} \right)}^2}},
\ee
as well as yields 
\be\label{en}
H=\frac{m c^2}{\sqrt{1-{\left(\frac{v}{c} \right)}^2}}=m c^2 \cosh{\frac{\theta}{c}}
\ee
for the energy of a free relativistic particle. The rightmost expression in (\ref{en}) 
is the Ruijsenaars--Schneider Hamiltonian for a single relativistic particle moving on a real line. Particle's momentum 
$P$ is found from the relativistic mass--shell condition
\be\label{mom}
{\left(\frac{H}{c} \right)}^2-P^2=m^2 c^2 \quad \rightarrow \quad P=m c \sinh{\frac{\theta}{c}}.
\ee

In order to construct a many--body interacting generalization, one invokes symmetry argument.
Considering $\theta$ in (\ref{tr1}), (\ref{vp}) to be an infinitesimal parameter
\be 
t'=t+\frac{\theta}{c^2} x, \qquad x'=x+\theta t,
\nonumber
\ee
one finds the generator of the Lorentz transformation 
$K=\frac{1}{c^2} x \partial_t+ t \partial_x$ 
which jointly with the generators of the temporal and spatial translations $H=\partial_t$ and 
$P=\partial_x$ obeys the structure relations of the Poincar\'e algebra in two spacetime dimensions
\begin{align}\label{str}
[H,P]=0, && [H,K]=P, && [P,K]=\frac{1}{c^2} H.
\end{align}

Eqs. (\ref{en}), (\ref{mom}) and (\ref{str}) underlie 
the Ruijsenaars--Schneider construction \cite{RS,RS1} which is implemented within
the Hamiltonian framework. Firstly, one notices that the functions
\begin{align}\label{free}
&
H=m c^2 \sum_{i=1}^N \cosh{\left(\frac{p_i}{mc} \right)}, 
&& P=m c \sum_{i=1}^N \sinh{\left(\frac{p_i}{mc} \right)}, && K=-m \sum_{i=1}^N x_i
\end{align}
defined on a phase space parametrized by the canonical pairs $(x_i,p_i)$, $i=1,\dots,N$,
obey the algebra (\ref{str}) under the conventional Poisson bracket $\{x_i,p_j \}=\delta_{ij}$. 
Then one alters the first two expressions in (\ref{free}) by introducing the interaction potential
\be\label{int}
H=m c^2 \sum_{i=1}^N \cosh{\left(\frac{p_i}{mc} \right)} \prod_{k\ne i} f(x_i-x_k), \quad
P=m c \sum_{i=1}^N \sinh{\left(\frac{p_i}{mc} \right)}\prod_{k\ne i} f(x_i-x_k), 
\ee
with $f(x)$ to be fixed below, keeps $K$ intact, and finally demands the structure relations (\ref{str}) to be satisfied
under the Poisson bracket.

The brackets involving $K$ hold automatically, while $\{H,P\}=0$ forces one to assume that $f(x)$ is 
an even function of its argument which additionally obeys the functional equation \cite{RS}
\be\label{FE}
\sum_{i=1}^N \partial_i \prod_{j \ne i} f^2 (x_i-x_j) =0,
\ee
where $\partial_i=\frac{\partial}{\partial x_i}$.

Taking into account the nonrelativistic limit 
\be
\lim_{c \to \infty} (H_{rel}-m c^2 N)=H_{nr},
\nonumber
\ee
and appealing to earlier results by Calogero \cite{C} and Sutherland \cite{S} on integrable nonrelativistic many--body systems, 
one can construct three explicit solutions 
to the functional equation (\ref{FE})
\bea\label{pot}
&&
f_r (x)={\left(1+\frac{g^2}{m^2 c^2 x^2}\right)}^{\frac 12}, \qquad
f_{tr}(x)={\left(1+\frac{\sinh^2 {\left(\frac{\nu g}{ mc} \right)}}
{\sin^2 {\left(\nu x \right)}} \right)}^{\frac 12}, 
\nonumber\\[2pt]
&&
f_h(x)={\left(1+\frac{\sin^2 {\left(\frac{\nu g}{ mc} \right)}}
{\sinh^2 {\left(\nu x \right)}}
\right)}^{\frac 12},
\eea
where $g$ and $\nu$ are arbitrary real parameters. They give rise to what is nowadays known as the rational, 
trigonometric, and hyperbolic Ruijsenaars--Schneider models \cite{RS}. 

\vspace{0.5cm}

\newpage
\noindent
{\bf 3. Rational Ruijsenaars--Schneider model with cosmological constant}\\

As is well known, the Poincar\'e algebra (\ref{str}) can be viewed as 
a contraction of the (anti) de Sitter algebra 
\begin{align}\label{ads}
[H,P]=\pm \frac{1}{R^2} K, && [H,K]=P, && [P,K]=\frac{1}{c^2} H,
\end{align}
in which the characteristic time $R$ goes to infinity \cite{BL,GP}. In physics literature, 
$\pm \frac{1}{c^2 R^2}$ is identified with the cosmological constant. 
Our objective in this section is to study a possibility to deform
the Ruijsenaars--Schneider models above so as to include a cosmological constant. For physical reasons 
(see Eq. (\ref{unst}) below), 
we focus on the anti de Sitter algebra (negative cosmological constant), which corresponds to the lower sign in 
$[H,P]$ above.

A natural starting point is to alter a free particle realization of the Poincar\'e algebra in the preceding section 
\be\label{start}
H=m c^2 F(x) \cosh{\frac{p}{m c}}, \qquad P=m c F(x) \sinh{\frac{p}{m c}}, \qquad K=-m x,
\ee
where $(x,p)$ is a canonical pair obeying the Poisson bracket $\{x,p \}=1$, and $F(x)$ is a function to be fixed 
below. 
Demanding the structure relations (\ref{ads}) to hold under the Poisson bracket, one finds the differential 
equation to fix $F(x)$
\be
c^2 R^2 F(x) F'(x)-x=0, \qquad \rightarrow \qquad F(x)=\pm \sqrt{\alpha+\frac{x^2}{c^2 R^2}},
\ee
where $\alpha$ is a constant of integration. 

Because the nonrelativistic limit of the (anti) de Sitter algebra results in the
Newton--Hooke algebra \cite{BL}, consistency requires the nonrelativistic limit of the 
Hamiltonian $H$ in (\ref{start}) 
to reproduce that of a particle moving in the Newton--Hooke spacetime \cite{GP}
\be\label{sup2}
\lim_{c \to \infty} (H-m c^2)=\frac{1}{2m} p^2+\frac{m}{2 R^2} x^2.
\ee
Eq. (\ref{sup2}) determines the form of the factor $F(x)$ entering (\ref{start})
\be
F(x)=\sqrt{1+\frac{x^2}{c^2 R^2}}.
\ee

Had one started with the de Sitter algebra involving $[H,P]=\frac{1}{R^2} K$ 
(positive cosmological constant), one would have obtained an unstable
mechanical system governed by the Hamiltonian
\be\label{unst}
H=\frac{1}{2m} p^2-\frac{m}{2 R^2} x^2
\ee
in the nonrelativistic limit. For physical reasons, this instance should be discarded.

Note that the second order equation of motion\footnote{Because (\ref{HO}) 
does not involve the speed of light, it maintains its form after implementing the nonrelativisitc limit. 
This correlates with the fact that the harmonic oscillator is known to 
be $SO(2,1)$ invariant system.}
\be\label{HO}
\ddot x+\frac{x}{R^2}=0,
\ee
which is equivalent to two first order canonical equations resulting from the Hamiltonian 
in (\ref{start}), 
can alternatively be obtained from the geodesic equations associated with the metric
\be\label{met}
\left(1+\frac{x^2}{c^2 R^2} \right) \left(c^2 dt^2-dx^2 \right).
\ee
The latter is prompted by the mass--shell condition
\be
{\left(\frac{H}{c F(x)}\right)}^2-{\left(\frac{P}{F(x)} \right)}^2=m^2 c^2.
\ee 

A single--particle pattern above, suggests the way of how to proceed when building an 
interacting system.
One starts with the ansatz
\bea\label{INT}
&&
H=m c^2 \sum_{i=1}^N \sqrt{1+\frac{x_i^2}{c^2 R^2}} \cosh{\left(\frac{p_i}{mc} \right)} 
\prod_{k\ne i} f(x_i-x_k), 
\nonumber\\[2pt]
&&
P=m c \sum_{i=1}^N \sqrt{1+\frac{x_i^2}{c^2 R^2}} \sinh{\left(\frac{p_i}{mc} \right)}\prod_{k\ne i} f(x_i-x_k), 
\nonumber\\[2pt]
&&
K=-m \sum_{i=1}^N x_i,
\eea
where $f(x)$ is assumed to be an even function of its argument, which is independent of the parameter $R$,
and requires the structure relations (\ref{ads}) to hold under the Poisson bracket. 

Like in the preceding section, all restrictions upon the form of $f(x)$ come 
from the bracket $\{H,P\}=-\frac{1}{R^2} K$. Collecting terms without the factor 
$\frac{1}{R^2}$, one reproduces the functional relation (\ref{FE}), while contributions involving $\frac{1}{R^2}$
yield
\be\label{cond}
\sum_{i=1}^N \left(2 x_i-\partial_i\left( x_i^2 \prod_{j \ne i} f^2 (x_i-x_j) \right) \right)=0.
\ee

Because the restriction (\ref{FE}) continues to hold, three prepotentials (\ref{pot}) 
can be checked against the additional condition (\ref{cond}), which is a consequence 
of the presence of a cosmological constant in the Lie algebra structure relations (\ref{ads}). 
It is straightforward to verify that only the rational model 
passes the hurdle.

Focusing on the rational model and analyzing the nonrelativisit limit 
$\lim_{c \to \infty} (H_{rel}-m c^2 N )=H_{nr}$, one reproduces the
Calogero model in the harmonic trap
\be\label{CM}
H_{nr}=\frac{1}{2m} \sum_{i=1}^N p_i^2+\sum_{i<j}^N \frac{g^2}{m {(x_i-x_j)}^2}
+\frac{m}{2 R^2}\sum_{i=1}^N x_i^2.
\ee
Within the nonrelativistic framework, the latter term correctly 
represents the universal cosmological attraction \cite{GP}.

To summarize, eqs. (\ref{INT}), in which $f(x)=f_r (x)$ and $f_r (x)$ is given in (\ref{pot}), 
determine a one--parameter deformation of the rational Ruijsenaars--Schneider system originating from
the anti de Sitter algebra (\ref{ads}). 

Concluding this section, it is worth mentioning that the anti de Sitter algebra 
in $1+1$ dimensions 
is isomorphic to the conformal algebra $so(2,1)$. Indeed, the linear 
change of the basis
\be
h=\frac{R}{c} \left(H+c P \right), \qquad d=c K, \qquad k= c R \left(H-c P \right),
\ee
brings (\ref{ads}) to the form
\be
[h,d]=h, \qquad [h,k]=2d, \qquad [d,k]=k, 
\ee
which coincides with the conventional $so(2,1)$ structure relations.
Hence, eqs. (\ref{INT}) can alternatively be viewed as providing a new 
dynamical realization of the conformal group $SO(2,1)$ within the framework of many--body 
mechanics in one dimension.

\vspace{0.5cm}

\noindent
{\bf 4. Equations of motion}\\

It is interesting to inquire how the presence of a cosmological constant affects the original 
equations of motion and particle dynamics. To this end, it proves convenient to represent the Hamiltonian in (\ref{INT})
in the manifestly positive--definite form 
\be
H=\sum_{i=1}^N \left(\lambda^{+}_i \lambda^{+}_i+ \lambda^{-}_i \lambda^{-}_i\right),
\ee
where 
\be\label{LF}
\lambda^{\pm}_i={\left(\frac{m c^2}{2} e^{\pm \frac{p_i}{m c}} \sqrt{1+\frac{x_i^2}{c^2 R^2}}
\prod_{k\ne i} f_r(x_i-x_k)  \right) }^{\frac 12},
\ee
with $f_r (x)$ in (\ref{pot}), and then compute the algebra of the subsidiary functions 
$\lambda^{\pm}_i$ 
\bea\label{ALF}
&&
\{\lambda^{+}_i,\lambda^{+}_j\}=\frac 12 W(x_i-x_j) \lambda^{+}_i \lambda^{+}_j, \qquad
\{\lambda^{-}_i,\lambda^{-}_j\}=-\frac 12 W(x_i-x_j) \lambda^{-}_i \lambda^{-}_j, 
\nonumber\\[2pt]
&&
\{\lambda^{+}_i,\lambda^{-}_j\}=-\frac 12 \lambda^{+}_i \lambda^{-}_i 
\left( \frac{1}{m c \left(1+\frac{x_i^2}{c^2 R^2} \right)} \frac{x_i}{c^2 R^2} 
-\sum_{k \ne i} W(x_i-x_k)\right) \delta_{ij},
\eea
where $\delta_{ij}$ is the Kronecker delta and $W(x_i-x_j)$ reads
\be\label{W}
W(x_i-x_j)=
\left\{
\begin{aligned}
0 \qquad \qquad ,&\qquad i=j\\
\frac{1}{m c \left( x_i-x_j\right) \left(1+\frac{{(m c)}^2 {\left( x_i-x_j\right)}^2}{g^2} \right)}
,&\qquad i\ne j.\\
\end{aligned}
\right.
\ee
Note that $W(x)$ is independent of the cosmological constant.

Taking into account the Poisson brackets
\be
\{x_i, \lambda^{\pm}_j \}=\pm \frac{1}{2 m c} \lambda^{\pm}_j  \delta_{ij},
\ee
one can easily obtain the Hamiltonian equations of motion which govern the evolution of $x_i$ and 
$\lambda^{\pm}_i$ over time
\bea\label{xl}
&&
{\dot x}_i=\frac{1}{m c} \left({\left(\lambda^{+}_i \right)}^2 -{\left(\lambda^{-}_i \right)}^2 \right),
\\[2pt]
&&
{\dot\lambda}^{+}_i=\sum_{j \ne i} W(x_i-x_j) \lambda^{+}_i {\left(\lambda^{+}_j \right)}^2
-\lambda^{+}_i {\left(\lambda^{-}_i \right)}^2 
\left( \frac{1}{m c \left(1+\frac{x_i^2}{c^2 R^2} \right)} \frac{x_i}{c^2 R^2} 
-\sum_{j \ne i} W(x_i-x_j)\right),
\nonumber
\eea
\bea
&&
{\dot\lambda}^{-}_i=-\sum_{j \ne i} W(x_i-x_j) \lambda^{-}_i {\left(\lambda^{-}_j \right)}^2
+\lambda^{-}_i {\left(\lambda^{+}_i \right)}^2 
\left( \frac{1}{m c \left(1+\frac{x_i^2}{c^2 R^2} \right)} \frac{x_i}{c^2 R^2} 
-\sum_{j \ne i} W(x_i-x_j)\right).
\nonumber
\eea

The first equation in (\ref{xl}) allows one to express momenta $p_i$ in terms of $x_i$ and ${\dot x}_i$
\be\label{exp}
e^{\pm\frac{p_i}{m c}}=\frac{\pm{\dot x}_i+y_i}{c \sqrt{1+\frac{x_i^2}{c^2 R^2}} 
\prod_{k\ne i} f_r (x_i-x_k)},
\ee
where we denoted
\be\label{y}
y_i=\sqrt{{\dot x}_i^2+c^2 \left(1+\frac{x_i^2}{c^2 R^2} \right) \prod_{k\ne i} f_r^2(x_i-x_k)},
\ee
while differentiating ${\dot x}_i$ with respect to the temporal variable and taking into 
account (\ref{xl}) and (\ref{exp}), one obtains the desired equations of motion
\bea\label{eom}
&&
{\ddot x}_i=m c \sum_{j \ne i} W(x_i-x_j) \left({\dot x}_i {\dot x}_j+ y_i y_j \right)
\nonumber\\[2pt]
&&
\quad \quad 
-\left(\frac{x_i}{R^2}- m c^3 \left(1+\frac{x_i^2}{c^2 R^2} \right) \sum_{j \ne i} W(x_i-x_j)  
\right) \prod_{k \ne i} f_r^2 (x_i-x_k),
\eea
with $y_i$ given in (\ref{y}). The terms involving $\frac{1}{R^2}$ comprise the 
difference with the original rational Ruijsenaars--Schneider model. 

\begin{figure}[ht]
\begin{center}
\resizebox{0.5\textwidth}{!}{%
\includegraphics{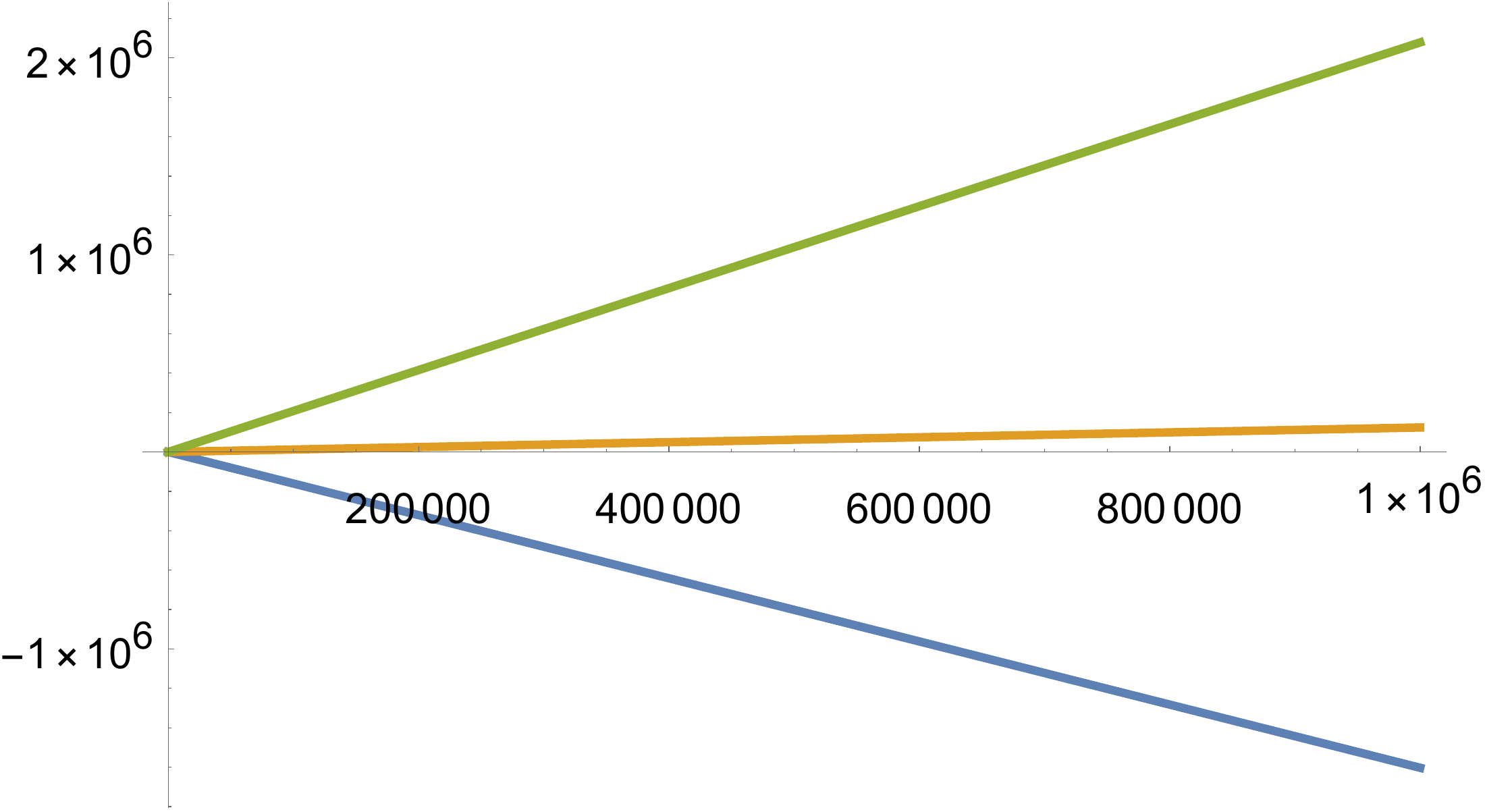}}\vskip-4mm
\caption{\small 
The graph of $x_i$ versus $t$ for the three--body rational Ruijsenaars--Schneider system 
with $m=c=g=1$, $x_1(0)=1$, ${\dot x}_1(0)=0.1$ (blue), $x_2(0)=2$, ${\dot x}_2(0)=0.2$ (orange), 
$x_3(0)=3$, ${\dot x}_3(0)=0.3$ (green) and $t \in [0,10^6]$.}
\label{fig1}
\end{center}
\end{figure}

\begin{figure}[ht]
\begin{center}
\resizebox{0.5\textwidth}{!}{%
\includegraphics{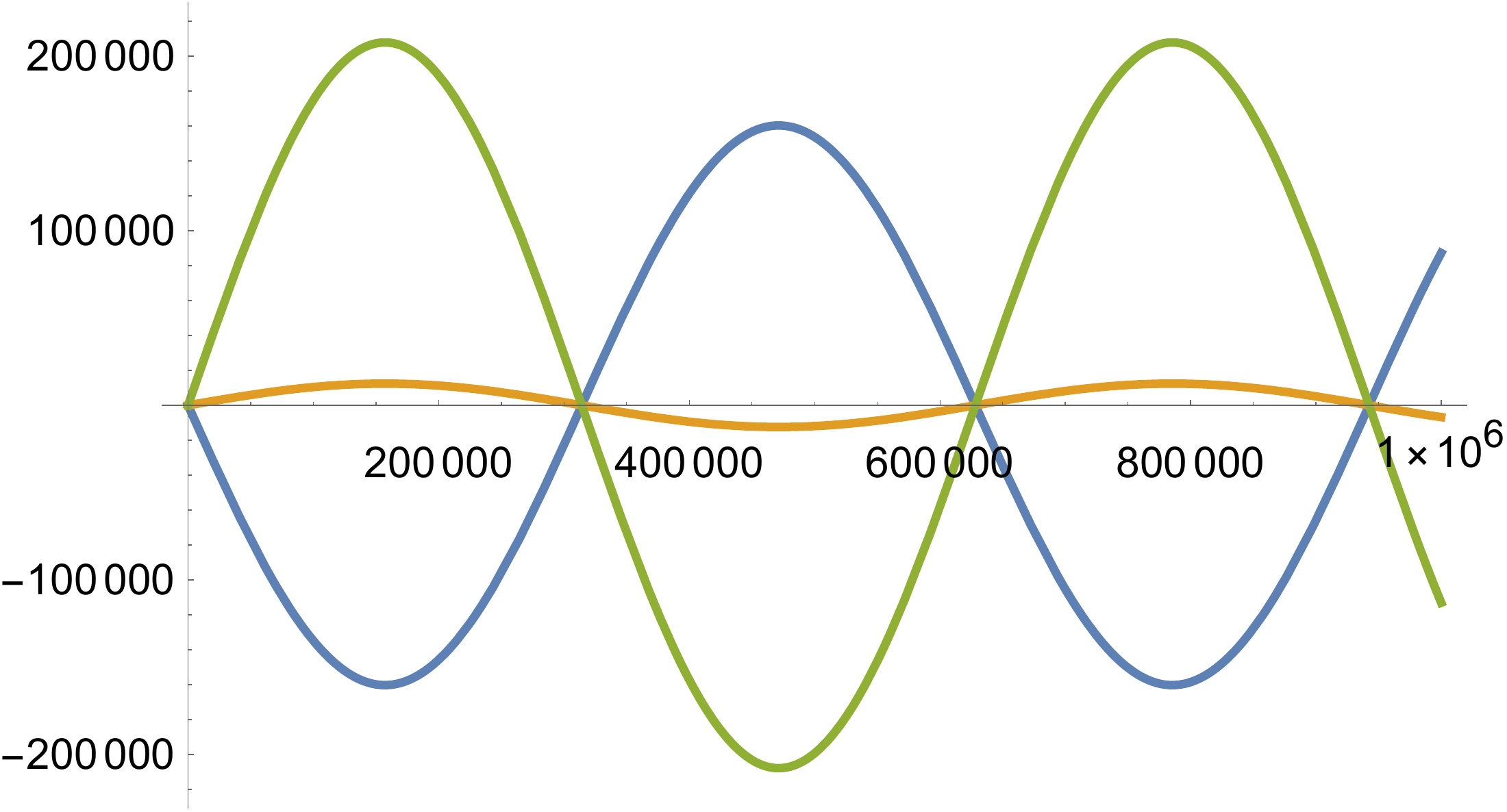}}\vskip-4mm
\caption{\small 
The graph of $x_i$ versus $t$ for the three--body rational Ruijsenaars--Schneider system with the cosmological constant
derived from $R=10^5$, and $m=c=g=1$, $x_1(0)=1$, ${\dot x}_1(0)=0.1$ (blue), $x_2(0)=2$, ${\dot x}_2(0)=0.2$ (orange), 
$x_3(0)=3$, ${\dot x}_3(0)=0.3$ (green), $t \in [0,10^6]$.}
\label{fig2}
\end{center}
\end{figure}

A qualitative difference in particle dynamics is illustrated in 
Fig. \ref{fig1} and Fig. \ref{fig2}. The former depicts a numerical solution of the equations of motion of 
the three--body model with vanishing cosmological constant for a particular choice of the free parameters 
$m=c=g=1$ and initial conditions $x_1(0)=1$, ${\dot x}_1(0)=0.1$ (blue), $x_2(0)=2$, ${\dot x}_2(0)=0.2$ (orange), 
$x_3(0)=3$, ${\dot x}_3(0)=0.3$ (green) on the time interval $t \in [0,10^6]$. The latter
displays solutions of (\ref{eom})  with $R=10^5$ for the same choise of the free parameters, initial conditions, 
and time interval. As is seen from the graphs, in the former case the motion is unbounded, while in the latter case 
the particles move along (quasi)periodic orbits.

Eqs. (\ref{eom}) reduce to
\be
{\ddot x}_i=\sum_{j\ne i} \frac{2 g^2}{m^2 {\left(x_i-x_j \right)}^3}-\frac{x_i}{R^2}
\ee
in the nonrelativistic limit $c \to \infty$,
which correctly reproduce the Lagrangian equations of motion associated with the Calogero model 
in the harmonic 
trap (\ref{CM}). Within the nonrelativistic framework, the Hooke's term $-\frac{x_i}{R^2}$ describes
the universal cosmological attraction \cite{GP}. As is seen in (\ref{eom}), its 
relativistic analogue is realized in a rather more fanciful way.

A salient feature of the Ruijsenaars--Schneider systems is that they are integrable. As 
explained in \cite{RS,RS1}, the integrability is maintained if one considers reduced models governed by 
either $H^{+}=\sum_{i=1}^N \lambda^{+}_i \lambda^{+}_i$ or 
$H^{-}=\sum_{i=1}^N \lambda^{-}_i \lambda^{-}_i$. 
In each case, the canonical equations 
of motion
\be
{\dot x}_i=\pm \frac{1}{m c}  {\left( \lambda^{\pm}_i \right)}^2, \qquad 
{\dot\lambda}^{\pm}_i=\pm \sum_{j \ne i} W(x_i-x_j) \lambda^{\pm}_i {\left( \lambda^{\pm}_j  \right)}^2
\ee
can be rewritten in the geodesic form
\be
{\ddot x}_i=2 m c  \sum_{j \ne i} W(x_i-x_j) {\dot x}_i {\dot x}_j,
\ee
with $W(x_i-x_j)$ exposed in (\ref{W}). Note that ${\dot x}_i>0$ for the dynamical system 
governed by $H^{+}$ meaning that particles are moving from left to right (right moving modes). 
For the model determined by $H^{-}$, they move in the opposite direction (left moving modes).

Although such models degenerate to free dynamics ${\ddot x}_i=0$ in the nonrelativistic limit,
they have been extensively studied in the past. Most notably, the corresponding equations of motion
can be put into the Lax form \cite{C}.
As follows from $W(x_i-x_j)$ in (\ref{W}), 
the explicit dependence on the cosmological constant is absent for the reduced systems, which 
correlates with the fact that in each respective case one has modes of one and the same type only,
their relative motion being insensible to the universal cosmological attraction.

\vspace{0.5cm}

\noindent
{\bf 5. The issue of integrability}\\

A salient feature of the Ruijsenaars--Schneider systems is that they are integrable. 
This is usually demonstrated by considering 
the Poisson--commuting set of functions \cite{RS}
\bea\label{FIRS}
&&
S^{+}_1=\sum_{i=1}^N e^{\frac{p_i}{m c}} \prod_{j \ne i} f(x_i-x_j), 
\\[2pt]
&&
S^{+}_2=\sum_{i<j}^N e^{\frac{p_i}{m c}+\frac{p_j}{m c}} \prod_{k \ne i,j} f(x_i-x_k) f(x_j-x_k), 
\nonumber\\[2pt]
&&
S^{+}_3=\sum_{i<j<k}^N e^{\frac{p_i}{m c}+\frac{p_j}{m c}+\frac{p_k}{m c}} 
\prod_{l \ne i,j,k} f(x_i-x_l) f(x_j-x_l) f(x_k-x_l), 
\nonumber\\[2pt]
&&
\dots
\nonumber
\eea
where $\dots$ designate higher order invariants of similar structure, 
verifying that $S^{-}_i$, $i=1,\dots,N$, which follow from
$S^{+}_i$ by reversing the sign of each $p_i$, can be algebraically built 
from $S^{+}_i$: $S^{-}_i=S^{+}_{N-i}/S^{+}_N$, with $S^{+}_{0}=1$,
and finally observing that the Hamiltonian in (\ref{int}) is a linear 
combination of $S^{+}_1$ and $S^{-}_1$. 

Because $S^{+}_i$ commute under the Poisson bracket, any member of the set can be chosen 
to define a Hamiltonian of an integrable system of the Ruijsenaars--Schneider 
type. In particular,
the Hamiltonians $S^{\pm}_1$ govern the dynamics of the right/left movers 
discussed in the preceding section. 

Introducing subsidiary functions similar to $\lambda^{+}_i$ 
in (\ref{LF})
\be\label{SF}
L^{+}_i={\left(e^{\frac{p_i}{m c}} \prod_{k\ne i} f_r(x_i-x_k)  \right) }^{\frac 12},
\ee 
one can rewrite (\ref{FIRS}) 
in the equivalent form 
\bea\label{IPL}
&&
S^{+}_1 =\sum_{i=1}^N {\left(L^{+}_i \right)}^2, \qquad 
S^{+}_2 =
\sum_{i<j}^N  {\left( L^{+}_i \right)}^2 
{\left( L^{+}_j \right)}^2 f^{-2}_r (x_i-x_j), 
\nonumber\\[2pt]
&&
S^{+}_3 =
\sum_{i<j<k}^N  {\left( L^{+}_i \right)}^2 {\left(L^{+}_j \right)}^2 
{\left( L^{+}_k \right)}^2 f^{-2}_r (x_i-x_j) 
f^{-2}_r (x_i-x_k) f^{-2}_r (x_j-x_k), 
\nonumber\\[2pt]
&&
\dots.
\eea
In this notation, the fact 
that the functions Poisson--commute is verified with the use of the relations
\bea\label{AR}
&&
\{L^{+}_i,L^{+}_j\}=\frac 12 W(x_i-x_j) L^{+}_i L^{+}_j,
\nonumber\\[2pt]
&&
\{L^{+}_i,f^{-2}_r (x_j-x_k) \}=-W(x_j-x_k) f^{-2}_r (x_j-x_k) L^{+}_i
\left(\delta_{ij}-\delta_{ik} \right),
\eea
where $W(x_j-x_k)$ is defined in (\ref{W}) and $\delta_{ij}$ is the Kronecker delta.

If a cosmological constant is present, a natural modification of (\ref{FIRS}) reads
\bea\label{FIRSCC}
&&
\mathcal{S}^{+}_1=\sum_{i=1}^N e^{\frac{p_i}{m c}} \sqrt{1+\frac{x_i^2}{c^2 R^2}} \prod_{j \ne i} f_r (x_i-x_j), 
\\[2pt]
&&
\mathcal{S}^{+}_2=\sum_{i<j}^N e^{\frac{p_i}{m c}+\frac{p_j}{m c}} 
\sqrt{\left(1+\frac{x_i^2}{c^2 R^2}\right) \left(1+\frac{x_j^2}{c^2 R^2} \right)}
\prod_{k \ne i,j} f_r(x_i-x_k) f_r(x_j-x_k), 
\nonumber\\[2pt]
&&
\mathcal{S}^{+}_3= \sum_{i<j<k}^N e^{\frac{p_i}{m c}+\frac{p_j}{m c}+\frac{p_k}{m c}} 
\sqrt{\left(1+\frac{x_i^2}{c^2 R^2}\right) \left(1+\frac{x_j^2}{c^2 R^2} \right) \left(1+\frac{x_k^2}{c^2 R^2} \right)}
\nonumber\\[2pt]
&& 
\qquad \times \prod_{l \ne i,j,k} f_r(x_i-x_l) f_r(x_j-x_l) f_r(x_k-x_l), 
\nonumber\\[2pt]
&&
\dots
\nonumber
\eea
while $\mathcal{S}^{-}_i$ are obtained by reversing the sign of each $p_i$. 
The fact that 
$\mathcal{S}^{+}_i$ form a Poisson--commuting set of functions can 
be established by deforming 
$L^{+}_i$ in (\ref{SF}) as follows
\be\label{SF1}
L^{+}_i \quad \rightarrow \quad
\mathcal{L}^{+}_i={\left(e^{\frac{p_i}{m c}} \sqrt{1+\frac{x_i^2}{c^2 R^2}} 
\prod_{k\ne i} f_r(x_i-x_k)  \right) }^{\frac 12},
\ee 
rewriting $\mathcal{S}^{+}_i$ in terms of $\mathcal{L}^{+}_i$ like in eqs. (\ref{IPL})
above, and finally verifying that $\mathcal{L}^{+}_i$ obey the same algebraic relations 
(\ref{AR}) as the original $L^{+}_i$.

In contrast to the flat case, $\mathcal{S}^{-}_i$ are no longer expressible 
in terms of $\mathcal{S}^{+}_i$, in particular
\be
\{\mathcal{S}^{+}_1,\mathcal{S}^{-}_1 \}=-\frac{2}{m c^3 R^2} \sum_{i=1}^N x_i,
\ee
which means that the above analysis of integrability 
should be modified accordingly. Rewriting the Hamiltonian in (\ref{INT}) in terms of 
$\mathcal{S}^{\pm}_1$
\be\label{HSS}
H=\frac{m c^2}{2} \left(\mathcal{S}^{+}_1+\mathcal{S}^{-}_1 \right),
\ee
and taking into account the Poisson brackets 
\be
\{\mathcal{S}^{\pm}_1, \sum_{i=1}^n x_i \}=\mp \frac{1}{m c} \mathcal{S}^{\pm}_1,
\ee
one can readily construct a conserved quantity
\be
I_1=\mathcal{S}^{+}_1 \mathcal{S}^{-}_1 
-\frac{1}{c^2 R^2} {\left( \sum_{i=1}^n x_i\right)}^2,
\ee
which links to the Casimir invariant of the anti de Sitter algebra (\ref{ads}).

Because the Hamiltonian (\ref{HSS}) is symmetric under the interchange 
of $\mathcal{S}^{+}_1$ and $\mathcal{S}^{-}_1$, it seems natural to search 
for other integrals of motion in the form 
\be
\mathcal{S}^{+}_2+\mathcal{S}^{-}_2+\dots, \qquad 
\mathcal{S}^{+}_3+\mathcal{S}^{-}_3+\dots, \qquad \mbox{etc.}
\ee
where $\dots$ designate extra contributions needed in order 
to ensure the commutativity with $H$.
It appears that the missing contributions can be built in terms of the
elementary monomials in $x_i$
\bea
M_1=\sum_{i=1}^N x_i, \qquad M_2=\sum_{i<j}^n x_i x_j, 
\qquad M_3=\sum_{i<j<k}^n x_i x_j x_k, 
\qquad \dots
\eea
In particular, a direct inspection of the three--body case reveals the following 
constants of motion
\be
I_2=\mathcal{S}^{+}_2+\mathcal{S}^{-}_2+\frac{2}{c^2 R^2} M_2,   \qquad
I_3=\mathcal{S}^{+}_3+\mathcal{S}^{-}_3-\frac{m}{c R^2} \{\mathcal{S}^{+}_1,M_3\}+
\frac{m}{c R^2} \{\mathcal{S}^{-}_1,M_3\},
\ee
where the last two terms are compactly written in terms of the Poisson brackets.

Curiously enough, the first integrals $I_1$, $I_2$, and $I_3$ do not commute with 
each other yielding higher order 
invariants.\footnote{The rational Ruijsenaars--Schneider model is known to be
superintegrable \cite{AF}. Hidden symmetries appear to be present for its counterpart 
with a cosmological constant as well. } This means that 
even more sophisticated analysis is needed in order to establish the 
Liouville integrability 
of the model at hand. In the latter regard, it is important to stress that the
Lax representation for the equations of motion without a cosmological constant 
is known in the literature for the right/left movers only \cite{C}. 
Were it available for the full system governed by the Hamiltonian 
in (\ref{int}), an amendment to include a cosmological constant 
would be rather straightforward. A transparent algebraic scheme enabling one to build 
Poisson commuting and functionally independent integrals of motion for the system under 
consideration remains a challenge.

\vspace{0.5cm}

\noindent
{\bf 6. Conclusion}\\

To summarize, in this work the original construction of Ruijsenaars and Schneider in \cite{RS}
was extended so as to include a cosmological constant. Specifically, starting 
with the anti de Sitter algebra (\ref{ads}), which is a one--parameter deformation 
of the Poincar\'e algebra in $1+1$ dimensions, and properly modifying 
the generators which furnish a dynamical realization, two functional equations were obtained, 
which determined possible interactions. The first equation coincided with that in \cite{RS}. 
The second restriction proved compatible with the rational model but ruled out
the trigonometric and hyperbolic variants. The issue of integrability was discussed in some detail. 
In particular, constants of motion characterizing the three--body case were explicitly constructed.

Turning to possible further developments, the most intriguing issue is to build a handy algebraic 
scheme of constructing Poisson--commuting and functionally independent integrals of motion for 
the system at hand. It appears that one has to reconsider the issue of representing the equations of motion
of the complete rational Ruijsenaars--Schneider model in flat space 
(not just the right or left movers) in the Lax form. In the latter regard, 
a link to reductions of matrix models \cite{FGNR} seems to provide a 
promising avenue. In particular, a matrix model progenitor of the 
rational Ruijsenaars--Schneider model with a cosmological
constant is worth studying. 

As is known, the rational Ruijsenaars-Schneider model together with 
its integrable structure can be obtained by the Hamiltonian reduction of the cotangent bundle to the
Lie group $GL(n,C)$ \cite{FK}.\footnote{The author thanks an anonymous JHEP reviewer for drawing his attention to this fact.} 
It is worth exploring a similar reduction
scheme for the extended model.

As was mentioned in sect. 3, the rational Ruijsenaars--Schneider model with a 
cosmological constant can alternatively be viewed as providing a new 
dynamical realization of the conformal group $SO(2,1)$ within the framework of many--body 
mechanics in one dimension. Conformal mechanics models of such a type have recently been
extensively studied in connection with near horizon black 
hole geometries (see e.g. \cite{AG2} and references therein). It would be interesting to understand 
whether the model in this work links to particles propagating 
on near horizon black hole backgrounds.

Our analysis in sect. 4 suggests that 
particles move along (quasi)periodic orbits. The construction of action--angle variables for the case 
at hand is an interesting open problem.

In a series of recent works \cite{AG1}, \cite{BDM}--\cite{AG4}, supersymmetric extensions of the
Ruijsenaars--Schneider models were built and their integrability was 
studied. It would be interesting to explore whether the analysis in this work 
is compatible with supersymmetry.

In a very recent work \cite{KKN}, Ruijsenaars--Schneider--type
models based upon twisted Poisson brackets, which are usually used to describe 
particles moving in an external magnetic field, have been proposed. A generalization
of the analysis in \cite{KKN} to the case of the anti de Sitter algebra 
is an interesting open problem.

\vspace{0.5cm}

\noindent{\bf Acknowledgements}\\

\noindent
This work was supported by the Russian Science Foundation, grant No 23-11-00002.

\end{document}